\documentstyle[natb_209]{mn}  

\def\dosingle#1::::{#1}  \def\dodouble#1::::{ } 

\dodouble \documentstyle[doublespacing,natb_209]{mn} ::::

\renewcommand\citep[1]{(\citealt{#1})}
\newcommand\citepf[1]{(\citealt*{#1})}    

\def\nice#1::::{#1}    \def\subm#1::::{}   

\newcommand\zzz[2]{#2}  

\def\S{section~}  

\def\k{\mbox{\rm\,km\,s$^{-1}$} }

\def\hMpc{\mbox{h$^{-1}$ Mpc}}
\def\hkpc{\mbox{h$^{-1}$ kpc}}
\def\hGyr{\mbox{h$^{-1}$ Gyr}}
\def\hMyr{\mbox{h$^{-1}$ Myr}}

\def\centreline{\centerline}
\def\.{{.}}
\def\gtapprox{\,\lower.6ex\hbox{$\buildrel >\over \sim$} \, }
\def\ltapprox{\,\lower.6ex\hbox{$\buildrel <\over \sim$} \, }
\def\sun{\odot}

\def\e{ {\scriptstyle \times} 10^}  

\def\arcs{\ifmmode {'' }\else $'' $\fi}     
\def\arcm{\ifmmode {' }\else $' $\fi}     
\def\deg{\ifmmode^\circ\else$^\circ$\fi}    
\def\ttimes{{\scriptstyle \times}}

\def\rComa{{\bmath{r}}_{\mbox{\rm \small Coma}}}  
\def\rinj{{r}_{\mbox{\rm \small inj}}}  
\def\ith{{i}^{\mbox{\rm \small th}}}  
\def\rtwod{{r}_{\mbox{\rm \small 2D}}}  

\def\xvec{\bmath{x}}
\def\xhat{\bmath{\hat{x}}}
\def\yvec{\bmath{y}}
\def\yhat{\bmath{\hat{y}}}
\def\vvec{\bmath{v}}
\def\nhat{\bmath{\hat{n}}}
\def\Av{\bmath{A}}
\def\Bv{\bmath{B}}
\def\Avp{\bmath{A}\bf{'}} 
\def\Bvp{\bmath{B}\bf{'}} 

\def\apj{ApJ}                 
\def\aj{AJ}                       
\def\aanda{A\&A}            
\def\cqg{ClassQuantGra}   
\def\mnras{MNRAS}

\def\frtoday{le\space\number\day\space\ifcase\month\or
  janvier\or f\'evrier\or mars\or avril\or mai\or juin\or
  juillet\or ao\^ut\or septembre\or octobre\or novembre\or d\'ecembre\fi\space \number\year}

\nice \input epsf ::::
\nice \def\mycaptionfont{\protect\footnotesize} ::::

\begin{document}

\newcommand\joref[5]{#1, #5, {#2, }{#3, } #4}
\newcommand\inpress[5]{#1, #5, #4}
\newcommand\epref[3]{#1, #3, #2}

\def\jref#1;;#2;;#3;;#4 {#1, {#2, }{#3, }#4}

\def\bref#1;;#2;;#3;;#4;;#5 {#1, {in #2 }(#3: #4), #5}
\def\apjpre#1;; {#1, preprint}

\def\apjpriv#1;; {#1, private communication}

\def\apjprpn#1;; {#1, in preparation}

\def\apjsub#1;;#2;; {#1, #2, submitted}

\def\apjprss#1;;#2;; {#1, #2, in press}

\def\CAMK{Nicolaus Copernicus Astronomical Center, 
ul. Bartycka 18, 00-716 Warsaw, Poland}
\def\IAP{Institut d'Astrophysique de Paris, 98bis Bd Arago, F-75.014 Paris,
France}
\def\IUCAA{Inter-University Centre for Astronomy and Astrophysics, 
Post Bag 4, Ganeshkhind, Pune, 411 007, India}

\title[Transverse Galaxy Velocities from Topology]{Transverse Galaxy Velocities from Multiple Topological Images}

\author[B.F.~Roukema \& S.~Bajtlik]{Boudewijn F.~Roukema$^{1,2,3}$ and Stanislaw Bajtlik$^1$\\ {$^1$\CAMK}\\ {$^2$\IAP} \\ {$^3$\IUCAA}}

\def\today{\frtoday}

\def\oo{$\ddot{\mbox{\rm o}}$}
\def\LaLu{Lachi\`eze-Rey \& Luminet (1995)}

\maketitle

\newfont{\sans}{cmss10}


\begin{abstract}
The study of the kinematics of galaxies within clusters or groups
has the limitation that only one of the three velocity components 
and only two of the three spatial components of a galaxy position
in six-dimensional phase space can normally be measured.
However, if multiple topological images of a cluster exist, then
the radial positions and sky plane mean velocities of galaxies 
in the cluster may also be measurable from photometry of the two
cluster images. 

The vector arithmetic and principles of the analysis are presented.
These are demonstrated by assuming the suggested topological identification
of the clusters RX~J1347.5-1145 and CL~09104+4109 to be correct and
deducing the sky-plane relative velocity component along the axis
common to both images of this would-be single cluster.

Three out of four of the inferred transverse velocities are
consistent with those expected in a rich cluster. 
A control sample of random `common' sky-plane axes, independent 
of the topological hypothesis,  
implies that this is not surprising. 
This shows that while galaxy kinematics are {\em deducible} from 
knowledge of
cosmological topology, it is not easy to use them to
{\em refute} a specific candidate manifold.
\end{abstract}


\def\tabone{
\begin{table*} 
\caption{\label{t-gals} Properties of matched galaxies: identity number; 
celestial positions in images (1) CL~09104+4109  and (2) RX~J1347.5-1145; 
apparent magnitude $R_2$ and the colour $C_2\equiv (B-R)_2$ 
in (2)
and their increments $\Delta R\equiv R_1-R_2,$  
$\Delta C \equiv (B-R)_1 - (B-R)_2$ (in mag);
transverse velocity component $v_{\perp}$ in direction $\nhat$ (in \k); 
average distance $r_\perp$ from centre in direction $\nhat$;
distances in background of centre $r_1$ and $r_2$ in images (1) and (2) 
respectively;
three-dimensional distance $r$from centre (in \hkpc). 
Notes: $^a$lower limit to $(B-R)_2$; $^b$not detected in $B$ in 
either image.}
$$\begin{array}{r cccc cccc rrrrr}
\hline \mbox{\rule[-0.3ex]{0cm}{2ex}}  
 \# & \alpha_1 & \delta_1 & \alpha_2 & \delta_2 &
 R_2 & C_2& \Delta R& \Delta C &
 v_{\perp}& r_{\perp}& r_1 & r_2 & r\\
\hline 
\mbox{\rule[-0ex]{0cm}{2ex}}  
   41 &     9^h 13^m 45.5^s &   40\deg 56\arcm 27.8\arcs &   13^h 47^m 30.6^s &  -11\deg 45\arcm  9.8\arcs &  
 19.0 &  1.6 &  -1.8 &    -0.2 &       0 &      0 &      0 &      0 &      0  \\
\mbox{\rule[-0ex]{0cm}{2ex}}  
   32 &     9^h 13^m 43.9^s &   40\deg 56\arcm 54.4\arcs &   13^h 47^m 31.9^s &  -11\deg 45\arcm 11.1\arcs &  
 19.5 &  2.4 &   0.3 &    -0.6 &     752 &     72 &    -68 &   -130 &    157  \\
\mbox{\rule[-0ex]{0cm}{2ex}}  
   33 &     9^h 13^m 39.7^s &   40\deg 56\arcm 51.2\arcs &   13^h 47^m 27.8^s &  -11\deg 45\arcm  2.5\arcs &  
 19.8 &  1.6 &  -0.9 & <  1.2 &    2866 &    -70 &    169 &   -335 &    405  \\
\mbox{\rule[-0ex]{0cm}{2ex}}  
   38 &     9^h 13^m 41.0^s &   40\deg 56\arcm 34.2\arcs &   13^h 47^m 28.0^s &  -11\deg 45\arcm 52.7\arcs &  
 20.1 &  2.7^a &  -0.9 &    - ^b &    6330 &   -176 &    -15 &   -235 &    294  \\
\mbox{\rule[-0ex]{0cm}{2ex}}  
   34 &     9^h 13^m 48.2^s &   40\deg 56\arcm 44.0\arcs &   13^h 47^m 34.2^s &  -11\deg 45\arcm 20.5\arcs &  
 19.8 &  0.4 &  -0.3 & <  1.8 &    -461 &    140 &   -221 &    107 &    296  \\
\hline 
\end{array} $$
\end{table*}
} 

\def\figgeomZ{ 
\begin{figure*} 
\nice \centreline{\epsfxsize=13cm
\zzz{\epsfbox[0 0 586 460]{"`gunzip -c skygeomZ.eps.gz"} }
{\epsfbox[0 0 586 460]{"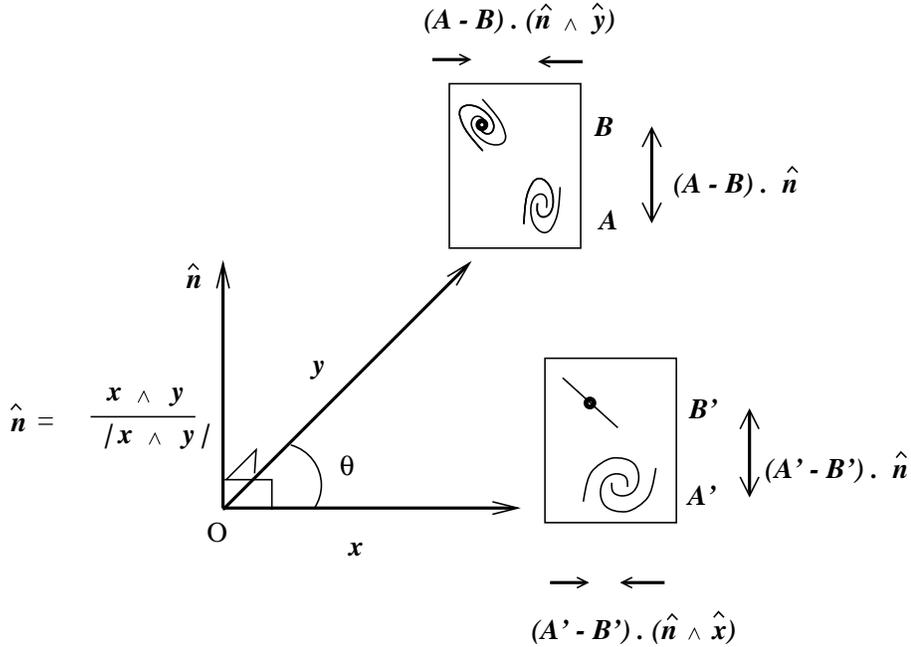"} }
}
 ::::
\subm \vspace{5cm} ::::
\caption{\label{f-geomZ} \mycaptionfont 
Relative positions of the observer (O) and 
pairs of galaxy images in 
two topological images of a cluster of galaxies, for a flat metric, 
and generators involving only translations (3-D plot). 
Schematic images of disk galaxies are indicated in order to 
show how these images would change when seen from different angles.
The two clusters are at vector positions 
$\xvec$ and $\yvec$, separated by some angle $\theta.$ 
The unit normal to $\xvec$ and $\yvec$
is $\nhat \equiv 
{ \xvec \wedge \yvec \over 
 | \xvec \wedge \yvec | }.$  The two galaxies 
are presumed to be at the mean distance of their parent cluster 
in each case, at vector positions  $\Av$ and  $\Bv$ in one cluster, 
and at  $\Avp$ and $\Bvp$ in the other cluster. 
 The separations of the two galaxies in the
$\nhat$ direction,  $(\Av-\Bv) .  \nhat$ and 
$(\Avp-\Bvp) .  \nhat,$ 
are measurable from the sky images in both cases, enabling a 
transverse velocity to be deduced.
}
\end{figure*} 
}

\def\figgeomXY{ 
\begin{figure*} 
\nice \centreline{\epsfxsize=13cm
\zzz{\epsfbox[0 0 586 476]{"`gunzip -c skygeoXY.eps.gz"} }
{\epsfbox[0 0 586 476]{"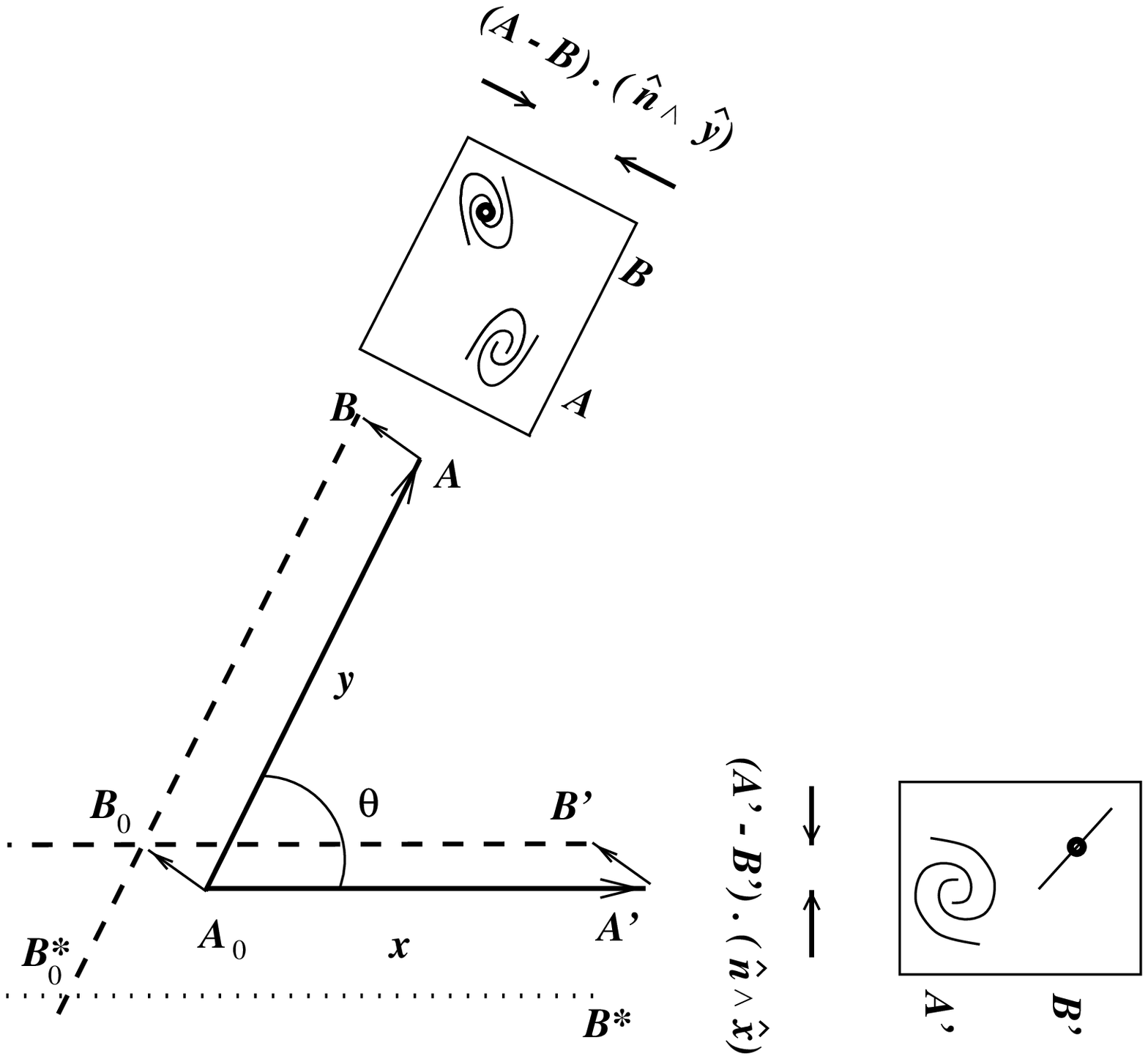"} }
}
 ::::
\subm \vspace{5cm} ::::
\caption{\label{f-geomXY} \mycaptionfont 
Spatial information available from pairs of galaxy images in two
topological images of a cluster, in the $\xvec - \yvec$ plane
(cf. Fig.~\protect\ref{f-geomZ}).  If the epochs of the two
topological images are identical, then the two vectors 
$(\Bv-\Av)$ and
$(\Bvp-\Avp)$ are identical, and their projections $(\Bv-\Av). (\nhat \wedge
\yhat)$ and $(\Bvp-\Avp). (\nhat \wedge \xhat)$ onto the two sky planes,
and in the $\xvec - \yvec$ plane, provide two linearly independent
components of $(\Bv-\Av)$ in the $\xvec - \yvec$ plane. These are
translated to the origin and combined to deduce the separation of $\Av$
and $\Bv$ in the $\xvec - \yvec$ plane: $(\Bv_0-\Av_0)$.
The point $\Bv^*$ is an alternative position for $\Bvp,$ showing
that a value of  $(\Bv^*_0-\Av_0)$ consistent with both sky images can be 
found even if the topological identification is incorrect. 
If the topological identification shown here were correct, then 
$\Bv$ would be slightly in the background of $\Av,$ and $\Bvp$ would
be clearly in the foreground of $\Avp.$
}
\end{figure*} 
}

\def\figBRx{ 
\begin{figure}
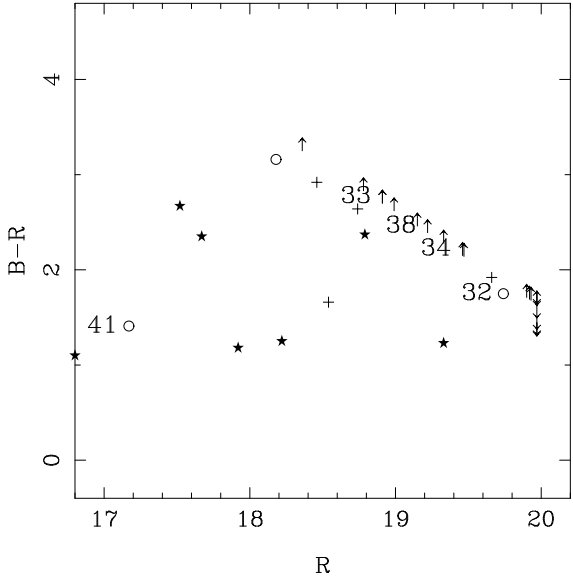
 
\nice \centreline{\epsfxsize=7.5cm
\zzz{\epsfbox[48 35 458 451]{"`gunzip -c BR09104.ps.gz"} }
{\epsfbox[48 35 458 451]{"BR09104.ps"} }
}
 ::::
\subm \vspace{5cm} ::::
\caption{\label{f-BR09104} \mycaptionfont 
$B-R$ colours and $R$ magnitudes of objects detected 
by APM software in  
digitised POSS plate at the position of CL~09104+4109 
(Fig.~\protect\ref{f-09104}). 
Objects having light
profiles in 
both bands characteristic of 
galaxies, stars or inconsistently as galaxies and stars 
are indicated as circles, stars and plus symbols respectively.
Galaxies undetected in the $B$ ($R$) band are shown as lower (upper) limits 
by arrows pointing up (down). The numbered objects are 
those satisfying the cluster membership criteria 
(\protect\S\ref{s-galsel}).
}
\end{figure} 
}

\def\figBRy{ 
\begin{figure}
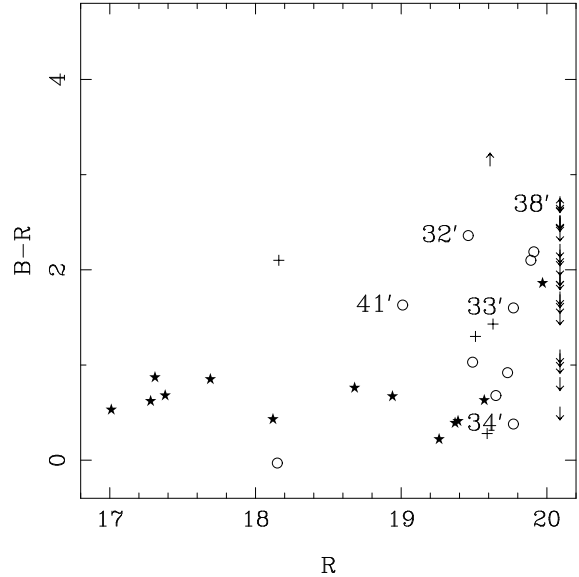
 
\nice \centreline{\epsfxsize=7.5cm
\zzz{\epsfbox[48 35 458 451]{"`gunzip -c BR1347.ps.gz"} }
{\epsfbox[48 35 458 451]{"BR1347.ps"} }
}
 ::::
\subm \vspace{5cm} ::::
\caption{\label{f-BR1347} \mycaptionfont 
$B-R$ colours and $R$ magnitudes of objects detected by APM software in 
digitised UK Schmidt survey plate 
at the position of RX~J1347.5-1145 (Fig.~\protect\ref{f-1347}). 
Symbols are as for Fig.~\protect\ref{f-BR09104}. As discussed in the
text, the blue galaxy $34'$ is also indicated.
}
\end{figure} 
}

\def\figskyx{
\begin{figure*}
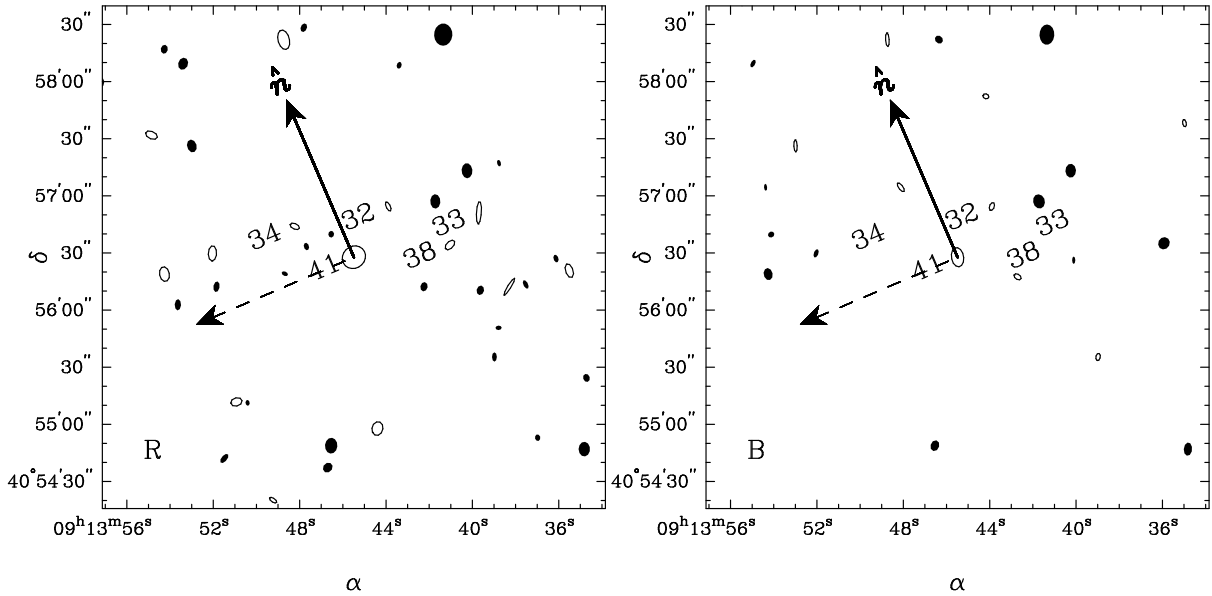
 
\nice \centreline{\epsfxsize=16cm
\zzz{\epsfbox[32 96 764 493]{"`gunzip -c sky09104.ps.gz"} }
{\epsfbox[32 96 764 493]{"sky09104.ps"} }
}
 ::::
\subm \vspace{5cm} ::::
\caption{\label{f-09104} \mycaptionfont 
APM scans of red and blue POSS plate photographic 
images of the core of the cluster CL~09104+4109.
East is left and North is up. Size is 
$4\.4\arcm \protect\ttimes 4\.4\arcmin.$ 
Hollow ellipses denote galaxy-like 
light profiles; solid ellipses star-like profiles. 
Ellipse sizes denote isophotal solid angles (not brightness).
Galaxies passing
the selection criteria discussed in 
\protect\S\ref{s-galsel} 
are numbered.
The vector shown in bold is $\protect\nhat,$ the normal common to both 
the lines-of-sight $\xvec$ (from the observer to this cluster) 
and $\yvec$ (from the observer to RX~J1347.5-1145). The other vector 
shown is $\nhat . \xhat.$ The two vectors are shown with modulus
$1\.5\protect\arcm$ $= 438$\hkpc. Galaxies 34, 33 and 38 are not detected
in the $B$ band. The cluster is at redshift $z=0\.442$ 
\protect\citep{Hall97}.
}
\end{figure*} 
}

\def\figskyy{
\begin{figure*}
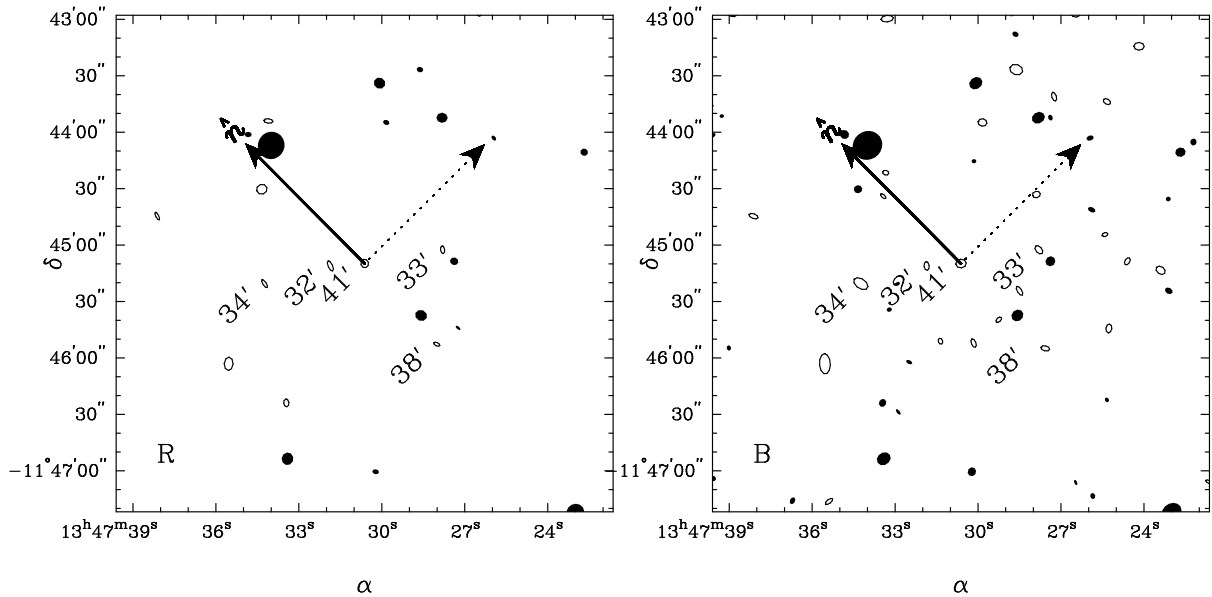
 
\nice \centreline{\epsfxsize=16cm
\zzz{\epsfbox[23 90 764 493]{"`gunzip -c sky1347.ps.gz"} }
{\epsfbox[23 90 764 493]{"sky1347.ps"} }
}
 ::::
\subm \vspace{5cm} ::::
\caption{\label{f-1347} \mycaptionfont 
Galaxies in the core of the cluster RX~J1347.5-1145, from 
UK Schmidt survey plate scans. Axes and symbols
are as for Fig.~\protect\ref{f-09104} except that primed numbers indicate
galaxies which would be identified with those in 
Fig.~\protect\ref{f-09104}, and the secondary vector is 
$\nhat.\yhat.$ The vector shown in bold is identical, 
in three-dimensional space, in both figures.
The two vectors are shown with modulus
$1\.5\protect\arcm$ $= 444$\hkpc. Galaxy 38{\arcm} is not detected in the 
$B$ band. The cluster is at redshift $z=0\.451$ 
\protect\citep{Sch96}, i.e. the image was emitted 
at a slightly earlier epoch than that of CL~09104+4109.
}
\end{figure*} 
}

\def\figvr{
\begin{figure} 
\nice \centreline{\epsfxsize=8cm
\zzz{\epsfbox[42 31 458 451]{"`gunzip -c v_r.ps.gz"} }
{\epsfbox[42 31 458 451]{"v_r.ps"} }
}
 ::::
\subm \vspace{5cm} ::::
\caption{\label{f-v_r} \mycaptionfont 
Resultant galaxy transverse velocities implied by topological identity
of the two clusters via a translation, shown in \k against 
the implied `two-dimensional' radial distances, in {\hkpc},
from the galaxy close to the centre of the cluster. 
The radial distances are defined as
$\rtwod \equiv |\; [ (A'-B'). (\nhat \wedge\xhat) ]\, (\nhat \wedge\xhat)
 + [(A'-B'). (\nhat \wedge\yhat) ]\, (\nhat \wedge\yhat) \;| $
 $\equiv \protect\sqrt{r^2- r_{\perp}^2},$ where $r$ and $r_{\perp}$ are 
defined as in Table~\protect\ref{t-gals}. 
That is, $\rtwod$ is the radial distance 
in the plane perpendicular to that of the
velocity component measured.
Solid circles indicate the implied velocities for the 
hypothesised common axis, 
other symbols indicate velocities calculated for random orientations 
of $\nhat$ in the two sky planes. 
The dotted lines indicate $\pm3$\,$\sigma_v,$ where $\sigma_v$ 
is a smooth fit to the velocity dispersions measured in
the Coma cluster by \protect\citet{KGunn82}.
}
\end{figure} 
}



\section{Introduction}
	Astronomical observations generally enable three elements 
of the position of an object in six-dimensional 
dynamical phase space to be determined:
two spatial elements by photometry and one velocity element (radial) by
spectroscopy. For extragalactic objects, various techniques enable
the cosmological expansion velocity to be approximately subtracted to deduce
local (`peculiar') radial velocities.

	In the study of the dynamics of a galaxy cluster, use of 
the mean redshift of the cluster implies that these three components
are relatively well determined. The other three components remain
undetermined, except when occasionally it can be argued that one galaxy
is in the foreground of another. To measure the transverse 
velocity of a galaxy at a redshift 
$0\.1 < z < 1$ to a precision of
100\k would require the detection of proper motions of around
$0\.01-0\.1 \mu$arcsec/yr.  
This is a signal about a thousand times more precise
than the noise (uncertainty) in 
typical VLBI estimates, e.g. of the motion of the Celestial Ephemeris
Pole \citep{Souch95}, so is not yet practical.

	The study of cluster dynamics, therefore, requires simplifying
statistical assumptions about the distribution of galaxies in 
phase space. While this is probably a reasonable approximation for
some purposes, measurement of all six elements of kinematical 
information for each galaxy would obviously enable a much more
detailed understanding of the cluster. For example, a net flow
of galaxies in a certain three-dimensional direction could be
compared to a cooling flow hypothesis or to a study of the merging
of sub-structure.

	The point of this paper is that measurement of the 
three missing kinematical parameters 
for galaxies in the centre of a cluster should be possible 
in certain cases of multiple topological imaging of
clusters. It would be possible to estimate 
both mean transverse velocities and line-of-sight relative galaxy
distances, simply by deep optical imaging.

	The reader is first briefly reminded of the nature
of multiple topological imaging. 

	In a standard Friedmann-Lema\^{\i}tre universe of constant
curvature, space (or more precisely, a hypersurface at constant
cosmological time) is a three-dimensional manifold of which both the
curvature and topology need to be measured in order to know its
geometry (e.g. \citealt{deSitt17,Lemait58}). The curvature can be
described by $\kappa_0\equiv \Omega_0+\lambda_0-1,$ where $\Omega_0$
is the density parameter and $\lambda_0$ is the dimensionless
cosmological constant. Together with the Hubble constant $H_0$ these
could be referred to collectively as the metric parameters.

	The metric parameters are related to local physics and so
are, in principle, easy to estimate or constrain by observation of
astrophysical objects. In practice, many observational complications
arise.

	The topology, which could require several parameters to
be fully described, is expected only to be related to global physics,
so strong theoretical predictions await developments in quantum
cosmology (e.g. \citealt{Hawking84,ZelG84}).
	
	So the principle of measuring topological parameters is 
purely observational, based on the fact that 
if the `size' of the Universe is smaller than the apparently 
observable sphere, then photons can travel several times `across' the 
Universe in less than the age of the Universe. In that case, 
astrophysical objects would be seen at different celestial positions
and different redshifts. The latter is equivalent to 
different distances and different cosmological epochs. These multiple images
are referred to as topological images. 

	Three-dimensional apparent space interpreted with the assumption
of a trivial topology would still be valid, and indeed very useful, 
to work with for many analyses, even though physically misleading. 
It would be tiled by `copies' of the Universe, and is
termed the `covering space'.

	Just as for techniques of estimating the metric parameters,
observational complications arise in searching for multiple topological
images. Indeed, the result of this paper suggests that galaxy kinematics 
in clusters are not likely to be useful in ruling out identity between
two images of clusters.

	The word `size' used above needs to be defined more precisely.
The size parameters used here are: the
`out-radius,' $r_+,$ which is the radius of the smallest sphere (in
the covering space) which totally includes the fundamental polyhedron; 
and the `injectivity radius', $\rinj,$ which is half of the
smallest distance from an object to any one of its topological images
\citepf{Corn98a}.
The terms `injectivity diameter' for $2\rinj$ 
and `out-diameter' for $2r_+$ are also adopted here.

	For reviews on cosmological topology, 
see \citeauthor{LaLu95} (1995; see also 
\citealt{Stark98,Lum98,LR99}), 
while recent developments include
theory of topology change at the quantum
epoch \citep{MadS97,Carl98,Ion98,DowG98,Rosal98,eCF98}, 
ideas for cosmological microwave background (CMB) 
methods \citepf{LevSS98,Corn98b,Weeks98}, a review of
three-dimensional methods \citep{RB98} and
observational analyses which include candidates for the 
topological parameters \citepf{RE97,BPS98}. 
See references in these
papers, \S\ref{s-cmb} here and \citet{Corn98a} 
for CMB-based arguments that $\rinj$ and $r_+$
either have or have not 
been constrained by the COBE satellite.

	The basic principle of measuring galaxy transverse velocities
is simple. Given the topology parameters to a certain precision, the
three-dimensional positions of multiple images of a galaxy known to
exist at a certain celestial position and distance (estimated by the
redshift) are calculated. If it is the case that several images of the
galaxy are expected to be separated by short time intervals, i.e.  at
similar redshifts, and at widely differing angles, then comparison of
optical images should be sufficient to estimate several mean
components of the galaxy's three-dimensional velocity over those time
intervals. In the case of two images separated by nearly a right
angle, a (near) transverse velocity can be estimated.

	The candidate topology\footnote{The word `topology' is 
used loosely here to mean `a 3-manifold of which some of the generators
are represented quantitatively in a common 
astronomical coordinate system'.} suggested by 
\citeauthor{BPS98} (1998, \S4.3)
to fit COBE data better than a `standard' CDM model, for
$\Omega_0=0\.8, \lambda_0=0\.0$, 
has a volume larger than that
of the observable sphere, so would not imply any multiple images of
ordinary astrophysical objects (it would only imply multiple partial images of
very large scale temperature fluctuations). 

	The candidate topological parameters which would be implied 
by the initial results of the quasar isometry search method of 
\citet{Rouk96} should imply multiple images well within the horizon
radius. However, the representations of negatively curved multi-connected 
manifolds are less simple than those of flat manifolds, 
and so would not be straightforward to apply.

	On the contrary, 
the candidate topology 
suggested by \citet{RE97}, according to which
the three rich clusters Coma, RX~J1347.5-1145 
\citep{Sch96} and CL~09104+4109 
\citep{Hall97}
would be three topological images of a single cluster, both implies
multiple topological images within the horizon and is simple to calculate,
since the angle formed by the three (with Coma at the vertex) is close
to 90\deg.
Moreover, it already includes the topological identification of three
known objects of which two are at nearly identical
epochs, and close enough that sky survey optical images are 
readily available. 

	Hence, the identification of these three clusters by the
translations (Coma $\rightarrow$ RX~J1347.5-1145) and 
(Coma $\rightarrow$ CL~09104+4109) in a flat ($\Omega_0=1$ or $\Omega_0=0\.2$,
and $\Omega_0+\lambda_0=1$) universe is adopted here for illustration
of the derivation of transverse galaxy velocities via multiple topological
imaging.

	As will be seen below, the results of this calculation show
that the converse is not easy: while topology can be used to deduce
galaxy kinematics, the expected kinematics of galaxies in clusters do
not provide an assumption-free constraint against this
cosmological topology candidate in the absence of a full scale
photometric and spectroscopic observing programme. 

In \S\ref{s-geom}, the geometry relating the clusters and cluster member
galaxies, the selection of galaxies hoped to be cluster members, 
and the matching of galaxies between two clusters 
are explained. In \S\ref{s-gal}, the application 
to digitised scans of photographic sky survey plates is presented 
and the resulting transversal velocities are deduced. 
In \S\ref{s-disc}, the results are discussed and observational 
arguments 
for and against the hypothesised topological identity of the two
clusters are listed. 
A summary is presented in \S\ref{s-concl}.

For reference, the reader should be reminded that 
the horizon diameter is $12000${\hMpc} for 
$\Omega_0=1$ ($\lambda_0=0$) and $\approx 23400 ${\hMpc} 
for $\Omega_0=0\.2$ ($\lambda_0=0\.8$). 
Except where otherwise stated, distances are quoted 
as comoving proper distances in 
the covering space of 
an $\Omega_0=1, \lambda_0=0$ universe 
(hereafter, `$\Omega=1$') or of an $\Omega_0=0\.2, \lambda_0=0\.8$ universe
(hereafter, `$\Lambda$')
and $h\equiv H_0/100$km~s$^{-1}$~Mpc$^{-1}$ is explicitly indicated.

\section{Method}
\label{s-geom}

The candidate topology of \citet{RE97} supposes that in two nearly 
perpendicular directions, 
the size of the Universe is $2\rinj \approx (965\pm5)${\hMpc} for $\Omega=1$ 
[$2\rinj\approx (1190\pm10)${\hMpc} for the $\Lambda$ model], and in
the third perpendicular direction the size is unknown, 
i.e. $r_+ > \rinj$. Because the
Universe has no boundaries, photons coming from greater distances 
in these directions would have in fact already crossed the Universe once
or more, so the rich clusters RX~J1347.5-1145 and CL~09104+4109 
would be images of the Coma cluster seen about $2\.8${\hGyr}
ago for $\Omega=1$ ($3\.3${\hGyr} for $\Lambda$). 
These two suggested topological images of Coma 
are seen at epochs separated
by only 35{\hMyr} for $\Omega=1$ (50{\hMyr} for $\Lambda$). 

If the topological hypothesis were correct, then this small time delay
would provide a relatively simple example of observationally deriving
transverse galaxy kinematics from identified clusters. Galaxy 
velocities in clusters are generally of order $\sim 1000$\k
$\approx 1$kpc~Myr$^{-1}$ $= 1$Mpc~Gyr$^{-1}$. Comparison of two images
of a cluster seen by a separation of 35{\hMyr} have galaxies 
which can move by only $\sim 35${\hkpc} over this interval,
so are unlikely to complete significant fractions of their
orbits between the two images.
On the contrary, two images of a cluster, even seen from the 
same direction, but separated by 3{\hGyr}, 
would show a complete scrambling of galaxy sky positions, making 
identification of corresponding galaxies difficult. 

So a criterion for simple determination of galaxy kinematics from
topology is that two topological images of a cluster be found which 
are separated by much less than a dynamical time.

In the case adopted here, this criterion is satisfied, and photographic
survey images are available for the two clusters. 

\subsection{Geometry}

\figgeomZ

\figgeomXY

In two images, 
there are two sky planes each containing two 
orthogonal directions, making four spatial axes available in total.
So, there is at least one spatial parameter which is known in both
images.  In the case of the hypothesised 3-manifold, in which the generators 
identifying topological images involve translations only (no rotations or
reflections), this spatial parameter can be easily deduced, as is shown
in Fig.~\ref{f-geomZ}.

For flat multi-connected models in which rotations are involved, 
or for negatively curved manifolds, the vectors representing the linear 
transformations from one topological image to another are not 
generally this simple, so derivation of the parameter in common 
would be more difficult. 

On the other hand, in the case of two images seen in nearly opposite directions
in the sky, again for translation-only generators, two spatial directions
would be common to both images. This would also apply to images seen
in nearly the same direction.

In the case under study here, in which $\theta \sim 90\deg,$ 
the four spatial axes can be chosen such that two are identical. This
common axis can be used to determine
transverse velocities in one sky direction. 

The remaining two axes can be used to fully determine the relative spatial
positions of a galaxy pair (Fig.~\ref{f-geomXY}),  
apart from relative 
movement between the two epochs of observation: 
\begin{equation} \label{e-spatial}
(\Bv_0-\Av_0) =
(\Bv-\Av)  + 
(\vvec_B-\vvec_A). (\nhat\wedge\xvec) \, \Delta t
\end{equation}
to first order in $\Delta t$.  This cannot be used as a consistency check
for the topological identity of the two images.
Any combination of two corresponding galaxy images 
at close projected distances from their respective cluster centres
implies a vector 
$(B_0-A_0)$ in the $\xvec- \yvec$ plane consistent with both images, even
if the galaxy images are not those of a single galaxy seen twice.

The accuracy of the determination of $(\Bv_0-\Av_0)$ 
depends on $\theta$. Values of
$\theta$ close to zero or $180\deg$ would make the uncertainty 
in the radial component of $(\Bv_0-\Av_0)$ high, but at the same lead 
to the transverse component being nearly similar in the two images,
so that both components of the transverse velocity can be calculated,
as mentioned above. An angle of $\theta=90\deg$ obviously minimises
the uncertainty if measurement of all three spatial separations is
desired. 

\subsection{Galaxy Identification} \label{s-galid}

How can objects in the two topological 
images of a cluster 
\begin{list}{(\alph{enumi})}{\usecounter{enumi}}
\item be identified as
galaxies which are members of the clusters and 
\item be matched as identical between the two images? 
\end{list}

Determination of cluster membership is not a new question 
(e.g. \citealt{KGunn82}, {\S}II.b; \citealt{Sarazin}, {\S}II.c).
If spectroscopy were available for candidate galaxies in both
images, then (a) galaxies which are very likely to be members of a
cluster can be decided to reasonable accuracy by a redshift
criterion. 
In that case, it would probably be better to be conservative 
in the decision on cluster membership, rather than maximally inclusive, 
in order to avoid wrong matches. That is,
galaxies which could ambiguously be interpreted either as 
high velocity cluster members or as projected 
foreground/background galaxies would be excluded from the sample.
The bias introduced in the distribution of transverse velocities
by this conservative inclusion criterion would be of less
immediate importance than that which would be introduced by 
a maximally inclusive criterion.

If deep CCD photometry were available for both images, 
then a possible technique for (b) would be to rank all galaxies
by apparent magnitude and colour and define a difference statistic
between the two ranked lists based on the differences in these
two parameters. To allow for (i) starbursts, (ii) supernovae, 
(iii) physical galaxy mergers, 
(iv) galaxies merged and/or hidden
by projection effects and (v) any other effects,
perturbations on the lists, in which a small fraction of 
galaxies can be dropped from either list, or shifted to very 
different positions, would need to be systematically generated
and the difference statistic again calculated.

The difference statistic should also include a parameter representing
inconsistency in the appearance of images for individual galaxies.
In Fig.~\ref{f-geomZ}, it is shown schematically that a disk galaxy
seen edge-on at $\Bvp$ may not appear as an edge-on galaxy at $\Bv.$

In fact, if the galaxy at $\Bvp$ is edge-on, i.e. at {90\deg} inclination,
and oriented exactly in the $(\Avp-\Bvp).\nhat$ direction, then 
at $\Bv$ the galaxy should be seen at an inclination of $90\deg-\theta,$
i.e. nearly face-on if $\theta\sim 90\deg.$
On the contrary, if the galaxy at $\Bvp$ is edge-on
and oriented perpendicularly to the $(\Avp-\Bvp).\nhat$ direction, then 
at $\Bv$ the galaxy should also be seen exactly edge-on. 
For deep photometric CCD images, 
a parameter for image inconsistency which combines all similar
possibilities should be combined with the difference statistic
representing magnitude and colour inconsistencies. It would obviously
be a much weaker constraint for ellipticals than for disk galaxies.

In order to unambiguously identify galaxies between the two lists,
the difference statistic would need to have a sharp minimum at the
correct perturbation. This would have to be modelled by starting
with a known list of galaxies in a cluster, making reasonable
assumptions for (i)-(v), taking into account the time difference
between the two images, and calculating the difference 
statistic for many realisations based on these assumptions.

Since the purpose here is only demonstration, and since the 
POSS photometry for CL~09104+4109 is not very deep 
(\S\ref{s-galsel}), the
criteria adopted here are simply :
\begin{list}{(\alph{enumi})}{\usecounter{enumi}}
\item objects with galaxy-like light profiles in the
$R$ band, 
within 400{\hkpc} of the cluster centre and having $B-R > 1\.5$
are considered to be cluster galaxies
(lower limits to $B$ magnitudes are adopted when the galaxy
is undetectable in $B$)
\item the galaxies are sorted by $R$ and identified between
the two images according to their rankings, allowing perturbations
for (i)-(v) above only if discussed explicitly.
\end{list}

$R$ band magnitudes are used here since they represent older
populations and so are more likely to be stable between the two
images.

The brightest galaxies in $R$ in the two images are defined for
the present purposes to be the cluster centres, and so considered
to be multiple topological images of a single galaxy. 
In fact, they
may not be exactly at the cluster centre (cf. \citealt{Lazz98}),
but since it is relative positions which are relevant here, this
seems the simplest robust assumption to make. Determination of
a `true' centre to the cluster in order to obtain a transversal
velocity of the `central' galaxy with respect to the true centre
could in principle be carried out by using centres determined
by gravitational lensing. For further applications of this method,
this would be an option to follow.

 In the present case, the lensing 
analysis of \citet{FTys97} seems to show that 
the brightest galaxy (in $R$) in the central region of RX~J1347-1145,
labelled galaxy \#41{\arcm} (Fig.~\ref{f-1347}),
is close to the gravitational centre of this cluster image. 
In addition, 
the brightest galaxy (in $R$) in the central region of CL~09104+4109,
labelled galaxy \#41 (Fig.~\ref{f-09104}), 
should correspond to the radio-quiet quasar 
which is also a strong IR source \citep{Hall97}. 
This quasar may be related to 
a cooling flow or a galaxy-galaxy merger, 
which would suggest that the galaxy should be close to
the centre. So the matching of these two galaxies (given the basic
hypothesis of identity of the clusters) seems reasonable.

The fact that galaxy \#41 lies slightly
outside of criterion (a) (since $B-R = 1\.4$) is unsurprising given 
the presence of the active galactic nucleus (AGN).
Strong blue emission from AGN is commonly observed, though 
the proportions between of light from young, blue, 
massive main sequence stars and non-stellar astrophysics are unclear.
The probability of the phenomenon just starting during
the time interval between the emission epochs of the two cluster images
is a factor considered in the identification of the two clusters. Since
this identification is the assumption made in order to carry out this
analysis, the matching of galaxies within the clusters should be made
consistently with this assumption.

So, the relative blueness of galaxy \#41 should not be used to exclude it,
and both criteria (a) and (b) need to be corrected in order to retain
consistency.

The correction adopted here is (a) to include this galaxy
since it would be more red without a starburst, and 
(b) if necessary, to slide its ranking in order that it matches 
the galaxy already labelled \#41\arcm.

\section{Results}
\label{s-gal}

Red and blue photographic images are available for 
CL~09104+4109 in the Palomar Optical Sky Survey (POSS) and for
RX~J1347.5-1145 in the UK Schmidt SRC Southern Sky Survey. 
Detections of objects in the digitisations of these plates by 
the APM (Automatic Plate Machine) group at Cambridge, UK, 
including morphological classifications into galaxy-like or
star-like light profiles were obtained. These are displayed
in Figs~\ref{f-09104} and  \ref{f-1347}, along with the
sky-plane orthonormal vectors discussed in \S\ref{s-geom}.

Since the POSS data is less deep that that of the UK Schmidt, 
the images scans of CL~09104+4109 are the limiting factor in comparing
the two clusters. This cluster is presented as being at 
the position of $\xvec$ and RX~J1347.5-1145 at the position of 
$\yvec.$ It should be remembered that the second of these two 
images is the cosmologically earlier of the two.

\subsection{Galaxy Selection} \label{s-galsel}

As mentioned above, CL~09104+4109 contains 
a hyper-luminous IRAS source.
Under the assumed hypothesis of topological identity of the
two clusters, the event causing this luminosity 
(e.g. a cooling flow or merger induced star-burst) would have started
during the 35{\hMyr} (50\hMyr) [for $\Omega=1$ ($\Lambda$) respectively] 
since the emission of the image of
RX~J1347.5-1145, and if explained by a star-burst would continue 
for a few hundred Myr longer.

This object is clearly visible 
as a very bright, moderately red object at the centre of Fig.~\ref{f-09104}. 

\figBRx

\figBRy

Figs~\ref{f-BR09104} and \ref{f-BR1347} show the colours and magnitudes
of the objects found in the two images, including numeric labels 
of galaxies as described in the following. 

Application of the selection criteria (a) 
(\S\ref{s-galid}) leads to five galaxies in 
CL~09104+4109 and four in RX~J1347.5-1145. The galaxies are ranked
by $R$ according to the suggested option for (b).
However, since five galaxies cannot be matched to four, 
we add to the second list the only other 
object within 400{\hkpc} of the centre of RX~J1347.5-1145 which is 
detected as a galaxy in both bands, galaxy \#34\arcm.
This galaxy would normally be excluded
from the criteria, so to retain physically consistent criteria, it
is necessary to suppose that the galaxy was undergoing a starburst
at the time of 
the RX~J1347.5-1145 image and that the young, blue stars had left
the main sequence by the time of the CL~09104+4109 image, in order that 
the galaxy was red enough to appear as one of the galaxies 
already included at this epoch. In that case,
the underlying $R$ magnitude of this galaxy without the starburst 
would be fainter than that measured, so this galaxy is added as 
the `faintest' galaxy around RX~J1347.5-1145 according to the
criteria (b).

As discussed above, perturbations in the ordering of the galaxies
should, in principle, be considered to take account of the 
many possibilities of luminosity evolution.
A single perturbation in the order is adopted here, i.e. that
of supposing galaxy \#32 around CL~09104+4109 to be one magnitude
brighter than measured.  This perturbation is adopted
in the numbering of identified galaxies in 
Figs~\ref{f-BR09104}, \ref{f-BR1347}, \ref{f-09104}, \ref{f-1347} 
and \ref{f-v_r}.

\figskyx

\figskyy

The image scans, shown in Figs~\ref{f-09104} and \ref{f-1347}, 
enable relative galaxy positions in the direction of the mutual
normal $\nhat$ to be judged by eye.


\figvr

\tabone

\subsection{Transverse velocities}
\label{s-prob}

Fig.~\ref{f-v_r} shows the resultant transverse velocities, 
plotted against a `two-dimensional' radial distance $\rtwod$ (see caption 
for definition), in order to make comparison with other observational
analyses as direct as possible.

For comparison, ten random axes in the two sky planes are generated 
and transverse velocities calculated just as for the hypothesised
axis $\nhat$ which is consistent between the two images. 

Both the hypothesised axis and the control axes give at least
one transverse
velocity which is obviously too high for a cluster member in each
case.  The results for the random axes shown by an `x' and by
a triangle are the most similar to those for the hypothesised axis.

In the case of the hypothesised axis, it is not possible to interpret
the high velocity as indicating {\em a} galaxy which is not a cluster 
member. It would indicate that the two matched galaxy images are
not of the same galaxy, and that at least one of these is probably a foreground
or background interloper. 
For galaxies seen in the lines-of-sight 
to two topologically identical clusters, only those close to 
the respective cluster distances would be common to both lines-of-sight
(apart from rare special cases). Interlopers would cause false matches,
and in the cases where the inferred velocity is unreasonably high, 
this would only indicate rejection of the galaxy-galaxy 
match, not a velocity of the interloper.

Table~\ref{t-gals} lists not only the transverse velocities but
also full three-dimensional information on the spatial displacements 
of galaxies from the cluster centre, as derived from 
equation~\ref{e-spatial}. Since $\theta\approx90\deg$ in the case
studied here, the three components are approximately orthogonal 
and $r^2 \approx r_\perp^2 + r_1^2 + r_2^2,$ but this would not be
the case in general. For example, for a topological image pair separated
by 45\deg the approximation would not hold.

If spectroscopic information were available for galaxies in the 
two images, then the remaining two velocity components would
also be measurable here.

The parameter $r_i$ (for the $\ith$ of the two images) 
indicates whether galaxies are in the foreground ($r_i <0$) or
the background ($r_i >0$) of the cluster centre.

\subsection{Uncertainties} \label{s-unc}

The positions generated by the APM are considered accurate to about
$0\.5${\arcs}, the colours to about $\pm0\.2$mag and the magnitude
zero-points to about $0\.25$mag (on-line facility). 
If we consider the uncertainties in the redshifts of the two cluster
images to be $\Delta(z)\sim 0\.001,$ i.e. 300\k, then 
the statistical
errors in the perpendicular velocities are about 17\%, dominated
by the uncertainty in the time interval implied by the 
uncertainties in the redshifts, while the foreground and background
offsets of galaxies are uncertain by 4{\hkpc}.

Since the uncertainty due to that in the time interval 
is essentially a fixed numerical value, it would become
fractionally smaller for 
an image pair separated by a larger time interval, making the transverse
velocity uncertainties smaller. However, a larger time interval would
also allow the possibility for galaxies to traverse larger fractions of
their orbits, so that the mean transversal velocity would become a less
useful quantity (e.g. a mean transversal, or radial, velocity over exactly one
orbit is zero). 

Whether the real Universe gives us a choice of enough cluster pairs
such that we have the luxury of choosing which pairs are best to be
analysed remains to be seen. The candidate examined here is all that
is available until further 
topological image pairs of clusters are proposed.

Systematic errors (apart from wrong matching of cluster pairs by 
cosmological topology)
are most likely to be present as the wrong matching
of galaxy image pairs. 
Wrong matches can be generated either by
the uncertainties in the zero-points and magnitudes 
of the galaxies which allow errors in the ranking procedure or by
non-cluster members which are included in the list for either cluster image.

In a full scale observing program to implement the method presented here,
CCD photometry and spectroscopy should increase the accuracy with
respect to all the different sources of uncertainties.

\section{Discussion} \label{s-disc}

Given the assumption of topological identity of two clusters
under a translation, it has been seen how galaxy transverse 
velocities can be deduced. 
Obviously, once (or if) the topology of the Universe is detected
by techniques such as the CMB technique of \citet{Corn98b} or 
the three-dimensional techniques of \citet{LLL96} 
or of \citet{Rouk96}, deep photometry of clusters would enable
methods such as those presented here to be applied in much more
detail to make a systematic study of transverse galaxy velocities
--- and full six-parameter kinematical information --- 
in a cluster.

In the present study, simple physical constraints based on sky-survey
image data result in quite reasonable transverse velocities for the
galaxies in the supposedly single cluster RX~J1347.5-1145/CL~09104+4109, 
apart from that of galaxy \#38. This latter `galaxy' would be best be
explained as a wrong galaxy-galaxy match, e.g. due to at least one
of the galaxies being an interloper.

\subsection{Photometric Evolution}
Are the photometric properties of the matched galaxies consistent 
with ordinary stellar evolution? The changes in apparent magnitude
and colour are listed in Table~\ref{t-gals} (expressed 
as the change in each parameter as cosmological time increases).

Apparent magnitude is related to absolute magnitude by the
luminosity distance, the K-correction (redshift effects for
a galaxy of non-evolving stellar population)
and the E-correction (evolutionary effect at a single redshift). 
The difference in luminosity distances of the two clusters implies 
only $0\.05$ brightening over the time interval and the difference
in K-corrections should also be small. 

If uncertainties $\delta(R)\approx \delta(B)$ are adopted, then 
the uncertainties for the photometry as cited in \S\ref{s-unc} 
imply  $\delta(R) \approx 0\.14$mag. Combining this
with the zero-point uncertainty gives 
$\delta(\Delta R) = 0\.41.$

Apart from galaxy \#41, which we know to be a strong IR source 
and therefore exceptional, and galaxy \#38, which should be
rejected on the ground of its high velocity, the 
remaining three galaxies are in fact consistent with 
no $R$ band evolution to $0\.7\sigma$ (galaxies 
\#32 and \#34) and to $2\.2\sigma$ (galaxy \#33). 
The latter galaxy would marginally have to brighten over the
time interval.

Because of the faintness of the galaxies, several are not detected
in $B$, particularly in the POSS plate, but the lower limits 
in $(B-R)$ are nevertheless useful for constraining colour
evolution. 

The uncertainty in the colour change is $\delta(\Delta C) = 0\.28$ 
(colour change is more precise than magnitude change due to the
zeropoint uncertainties in the magnitudes).
This implies inconsistency with no colour evolution 
at at least $4\sigma$ and $6\sigma$ for galaxies \#33 and \#34
respectively, in the sense that these galaxies become at
least $1\.2\pm0\.3$mag and $1\.8\pm0\.3$mag more red, 
respectively, over $35\pm5${\hMyr} ($\Omega=1$).

Both of these galaxies' colours in the earlier of the two
images, i.e. $C_2,$ are relatively blue. The normal explanation
for this is that they are seen at the end of a starburst. 

How rapidly can a starburst galaxy redden following the
end of a starburst?
As long as at least a few percent of the mass of a galaxy is 
involved in the starburst, the newly formed stars still dominate 
the optical spectrum up to $\sim100${Myr} after the starburst 
has terminated, so it is sufficient to consider the 
reddening of a stellar population created in an 
instantaneous starburst.
For example, see Fig.~4 of \citet{BC93}. 

Estimation from this 
figure indicates that (for a Salpeter initial mass function, IMF),
between 1 and 10~Myr from the occurrence of an instantaneous
starburst, the $(B-R)$ colour can redden by about $1\.4$mag.
On the other hand, between 1 and 10~Myr from the {\em onset} of
a constant star formation rate (SFR) period, the $(B-R)$ colour can
redden by about $0\.6$mag. At the same time, the $R$ band magnitudes
should brighten according to $\Delta R=-1\.1$ and $\Delta R=-3\.6$ 
respectively.

Given the uncertainties these 
values provide a fair guide for the time interval here of
$35\pm5${\hMyr}.

This would make galaxies  \#33 and \#34 consistent with 
the endpoints of starbursts to $0\.7 \sigma$ and
$1\.4 \sigma$ respectively, by their colours, and to 
$0\.5 \sigma$ and $2 \sigma$ respectively, in their 
$\Delta R$ values.

If reddening by dust were taken into account, many other 
possibilities would be possible.

The chance of seeing the end or the beginning of a starburst 
depends on whether the starbursts frequently seen 
in cluster galaxies at redshifts $z \gtapprox 0\.3$
\citepf{BO78,ODB97} are continuous and smooth or are composed of many
short star formation periods. Relatively many short bursts 
would be required for the probability of seeing the end 
or the begininning of a starburst to be significant.

In the case of a full observing program including 
spectrophotometry, detailed studies should enable stellar population
evolution of individual galaxies to become a check on
the consistency with the original hypothesis, but this is not
the case here.

\subsection{Further Work}

The determination of cluster membership and constraints on
stellar population evolution would obviously benefit from both
deep photometry and spectroscopy. 

In fact, the consistency of the kinematics and photometry for the
assumed topological image pair of clusters indicate that both
types of observations would be necessary if this method
were to be considered a means of 
refuting a cosmo-topological hypothesis. 

This would not necessarily be sufficient, however. Higher numbers of
fainter galaxies would enable more possible matches. Historical
experience with measurement of other cosmological parameters suggests
that these parameters are often more useful as inputs than
as outputs of small-scale galaxy studies, so the topological 
parameters do not seem to provide an exception.

So, although a full scale observing programme on these two cluster images
could be motivated by the desire to investigate how well this method
can exclude a candidate pair of topologically imaged clusters, it
could be more safely motivated by independent evidence strengthening
the candidate match.

\subsection{Arguments For and Against the Topological Identity}

For completeness, various arguments for and against the 
hypothesised image pair, 
and observational predictions based on the same hypothesis, 
are resumed below. The reader is reminded that the two clusters 
would be also physically identified with the Coma cluster.

\subsubsection{Likelihood of Finding Near Right-Angled Configurations}

This geometrical configuration in the covering space is
one which has already been searched for among other objects
(e.g. quasars, \citealt{Fag87}), and was found among a very small
number of objects by \citet{RE97}. An {\em a posteriori} 
calculation of the probability of finding this 
configuration, particularly given the subjective 
selection of the very small list of bright clusters, would be difficult
to do objectively.

The striking nature of the configuration
would not be a strong argument alone, but the history of literature
on hypertorus models (\S\ref{s-cmb})
and the simplicity of the configuration 
make it good as a {\em working hypothesis}, to see how tightly 
methods which attempt to constrain topology can really work if
considered with the same care as methods to constrain the metric
parameters, $\Omega_0, H_0$ and $\lambda_0.$

\subsubsection{CMB Analyses} \label{s-cmb}

To test the consistency of a candidate 3-manifold with CMB data would 
require application of the `identified circles' principle \citep{Corn98b}.

The identified circles principle is simply
the property that the sets of multiply imaged points on the SLS
would form pairs of identified circles, if non-trivial topology
were detectable. If locally anisotropic radiation from the SLS
(e.g. the Doppler effect) and foregrounds are insignificant,
then any specific candidate can therefore be
falsified by showing that 
the temperature fluctuations running around 
hypothetically identified circles are significantly different.
This is independent from assumptions regarding the statistics
of density perturbations, but does depend on the assumption that
the radiation is locally isotropic.

Because of the poor signal-to-noise ratio and resolution, 
this has not yet been systematically applied to COBE data to constrain 
multi-connected universe models. It is intended to 
be applied to observational data from the MAP and Planck satellites
(\citealt{Corn98a}). 

However, in spite of the poor signal-to-noise and resolution, several
CMB analyses based on COBE data have been attempted for hypertoroidal
universes (see \citealt{Corn98a} and references therein), by (i)
modelling individual 3-manifolds and (ii) assuming properties of the
perturbation spectrum --- Gaussian distributions in the real and
imaginary parts of the amplitudes of density perturbations and a power
spectrum of shape $P(k)\propto k^1$.  The authors claim that the COBE
data are inconsistent with hypertoroidal universes of $2\rinj
\ltapprox R_H/2$. Since in the present case $2\rinj \approx 0\.16 R_H$
for $\Omega=1$ (for $\Lambda$ we have $2\rinj \approx 0\.10 R_H$),
this would seem to provide evidence against the hypothesised
topological identity adopted for the illustration in this paper.

However, these analyses consist more of a test of the assumptions
(ii) on large scales rather than of the topology of space. 
They do not refute the hypertoroidal models tested: this would
require an application of the identified circles principle.

The theoretical motivation for (ii) is unlikely to be valid at scales
approaching $\rinj$ and $r_+$. That is, either for a hyperbolic or for a
flat, $\lambda_0 \sim 0.7$ metric to be presently observable, 
inflation needs some degree
of fine-tuning (which can partly be provided by the ergodicity of
geodesics in the former case, \citealt{Corn96}). Since curvature
or $\lambda_0 > 0$ must remain ``uninflated'' in the sense of 
being observable at the present epoch in these cases, 
it is unclear why perturbations on the scale of $R_C \sim R_H$ 
should necessarily be ``inflated'' in the sense of exactly satisfying
the assumptions on the power spectrum. Moreover, even for other
choices of metric, if the Universe has observable non-trivial topology,
then scales approaching $\rinj$ and $r_+$ 
need not necessarily be ``inflated'' either.

The observational motivation for (ii) is equally lacking for tests 
of non-trivial topology. 
The only observational justification of these properties
on large scales is COBE data, which is analysed 
{\em assuming trivial topology}. This cannot be used to test
non-trivial topology.

So the statistical confidence levels quoted for these papers
rejecting hypertoroidal universes should be interpreted as
statements about theoretical random realisations where 
the properties of large length scale ($r = 2\pi/k \sim \rinj < r_+$)
density perturbations are
{\em assumed} to be unaffected by the finiteness of the Universe
apart from a sharp cutoff.
(Note also that some authors find evidence for non-Gaussian distributions: 
\citealt*{Ferr98,Pando98}).

Nevertheless, even if it is assumed that 
the perturbation properties satisfy ``inflationary'' conditions
at the topology length scales, if one
disregards the other simplifications of the physics 
at the SLS in the standard COBE analyses, the possible contamination
by foregrounds, and the poor signal-to-noise 
and resolution ($\sim$ 10\deg) of the COBE maps, the present
hypothesis would still not be excluded.
The $T^2$ candidate suggested has 
$\rinj \ll R_H$ but $r_+ \gg R_H$. Since most work deals with $r_+,$
it does not apply in the present case.

Specifically, the orientation of the long (unconstrained) axis of the
toroidal universe hypothesised here is about 23{\deg} from the galactic
plane, so that the smallest cross-sections of copies of the
fundamental polyhedron with the surface
of last scattering (SLS) are mostly obscured by or at least
affected by galactic contamination.  The most visible cross-sections
are large. This would generate the large scale power, which should, 
in some average sense across different realisations of the universe,
cut off above a scale related to $r_+.$

In the work most relevant to the present case, \citet*{deOliv96} 
present an interesting method, where 
$\rinj < R_H \ll r_+$ hypertori are considered.
Fig.~1 (top right) of \citeauthor{deOliv96} 
should be similar to a CMB map expected for the $T^2$ candidate used
here, though it should also have the plane of the Galaxy removed, the long axis
oriented at about {23\deg} from the galactic plane rather than 90\deg,
and a `cell size' smaller by a factor of two. Fig.~3 of 
\citeauthor{deOliv96} would imply that our $T^2$ candidate is 
rejected at more than $2\sigma$ by the authors' $S_0$ statistic,
so this is potentially interesting.

However, apart from the fact that the $2\sigma$ rejection is 
a statement about the statistics of the perturbation simulations rather
than about observational inconsistency, the fact that the method
involves detecting symmetry implies that it is specially sensitive
to any asymmetries in the noise contributions.
Galactic foregrounds are far from being isotropic. 
So, a parameter representing symmetry in the CMB from the
component at the SLS,
due to the $T^2$ topology, strongly risks damping by asymmetry from
foregrounds.

If \citeauthor{deOliv96}'s method were to be applied more thoroughly,
in order to provide more serious constraints, it would be necessary to
consider 
(i) a wider range of assumptions on the fluctuation statistics, 
(ii) the integrated Sachs-Wolfe effect and the 
Doppler effect,
(iii) correlations in the galactic and extragalactic foregrounds,
and (iv) removal of the contaminated areas suggested by 
\citet{CaySm95}.

\subsubsection{Coincidences of Large Scale Structure}
\citet{RE97} noted that with the correct topological hypothesis, 
the distribution of large scale structure in walls, bubbles and filaments 
at scales around 
$50-150${\hMpc} (e.g. \citealt*{deLapp86,GH89,daCosta93,Deng96,Einasto97}), 
which quasars can be expected to
trace, should enable significant correlations to be measured. 
No significant correlation was found with the parameters published
in that paper. 

Since the auto-correlation function of galaxies
at $2\ltapprox z\ltapprox 3$ (where the majority of observed quasars
lie), may cover a range of amplitudes overlapping the present
values (e.g. \citealt{RVMB98} at small separations), this is
potentially a good method of refuting or strengthening a candidate
for the 3-manifold of the Universe. The method is in fact equivalent
to that of \citet{LLL96}, except that (a) individual objects are replaced
by `pixels' of filaments and walls in large scale structure 
of thickness not much less than 10\% of their separations, and 
(b) the `pixels' are sampled very sparsely and non-uniformly, 
so that the spikes expected would be much less strong than
in the plots of these authors.

Indeed, the $\sim 8000$ quasars presently identified with $z>1$ 
are both unevenly distributed and 
rare, with a mean separation of $\gtapprox 100${\hMpc}. This 
implies typically one or a few quasars per entire `unit' of large
scale structure, so that pairs of quasars seen in topologically
identified `pixels' would be rare.

Other factors are 
(a) it could be the case that the angles of the fundamental cube are not
perfect right angles, particularly since the identity between the 
three clusters is not exactly a right angle, 
and that the sides are of slightly unequal lengths;
and (b) the test should 
also be performed for a range of $\lambda_0$ values 
giving flat metrics.
It would be sufficient to be one degree in error in the orientation
or to have $\lambda_0$ mis-estimated by $0\.02$
to have a transversal error of $\sim 60$ {\hMpc} at typical quasar 
distances of $z\sim 3.$ This error is at the same scale as that of 
large scale structure, so would be likely to swamp any genuine 
signal.

This technique would therefore require at least an order of
magnitude more $z \gtapprox 1$ objects and 
detailed consideration of sources of uncertainty
in order to significantly 
refute candidates for the topology of the Universe, but in principle could
be powerful.

\subsubsection{Mass Estimates}
The total masses are identical within the uncertainties. 
\citet{Briel92} ROSAT estimates for Coma 
are $(4-10)\e{14}M_{\sun}$ to $1h_{50}^{-1}$~Mpc
and $(10-20)\e{14}M_{\sun}$ to $3h_{50}^{-1}$~Mpc; \citet{Sch96} give
$5\.8\e{14}M_{\sun}$ and $17\e{14}M_{\sun}$ for 
RX~J1347.5-1145 to the same radii respectively;
and the gas fraction to a large radius appears lower in Coma, but
consistent within the uncertainty. 

From gravitational lensing data, \citet{FTys97} infer the 
mass of RX~J1347.5-1145 as $(11\pm4)\e{14} M_\sun$
within the central $2h_{50}^{-1}$~Mpc, consistent with the above estimates
for a simple $M\propto R$ 
profile\footnote{The value cited in the abstract of 
\protect\citet{FTys97} appears to be a typographical error; 
see \S6.7 (p23) of that paper. Also note that 
\protect\citeauthor{Sch96}'s (1996) $M(r<1Mpc)$ value quoted
in these authors' Table~2 is for $H_0=50$km~s$^{-1}$~Mpc$^{-1},$
not $H_0=100$km~s$^{-1}$~Mpc$^{-1},$ so the last sentence in the
abstract of \protect\citeauthor{FTys97} also appears to be 
an error.}.

\subsubsection{X-ray Fluxes}

The X-ray flux and core ($r<500h_{50}^{-1}$~kpc) 
surface brightness of RX~J1347.5-1145 are considerably 
higher than those of Coma, 
e.g. $L_X(0\.1-2\.4$~keV$)$ of RX~J1347.5-1145 
is about ten times as high as that of Coma
\footnote{For $\Omega_0=1, \lambda_0=0, h=0\.5$.}.
Taking into account the X-ray `K-correction' would only 
make the intrinsic $L_X(0\.1-2\.4$~keV$)$ a small factor smaller.
For RX~J1347.5-1145 and Coma to be images
of one another, a large reduction in emission due to the
removal of a cooling flow, possibly due to the merger between
sub-clusters, would have to have occurred.

In fact, Coma does not have a cooling flow, and 
\citet{Biv96} argue that Coma has had a recent merger of 
two subclusters (and is ongoing a merger with another subcluster), 
so any cooling flows are very likely to have been disrupted.
Models by \citet{AllFab98} and \citet{Peres98} show that up to 70\%
of a cluster's luminosity can be due to the cooling flow,
so once this is removed, the luminosity of 
RX~J1347.5-1145/CL~09104+4109 could drop to close to that of Coma.
In addition, Table~1 and Fig.~2 of \citep{AllFab98} show that
bolometric luminosity estimates for these two images (which are 
at nearly the same redshift, so have little relative 
K-correction) agree within $1\.4 \sigma.$

\subsubsection{Coincidence of Seeing Galaxy \#41 Just Before and During
a Highly Luminous IR Phase}

As mentioned above, for 
RX~J1347.5-1145 and CL~09104+4109 to be identical, the quasar  
(lying in the galaxy labelled here as \#41) 
seen only in the latter 
would have to have started during the time interval separating the
two images. 

The part due to stellar luminosity,
mostly reradiated in the far-IR by dust but with some blue light escaping, 
would have a time scale for the massive stars to leave the main sequence 
of $\sim 100$~Myr. The time scale for the duration of this episode 
of AGN activity, if due to the merger of typical large galaxies
of $10^{12}M_{\sun},$ would be about $200$~Myr, while
\citet{Cavpad88} estimate that quasar lifetimes
are likely to be at most about a 1~Gyr. 
So if RX~J1347.5-1145 and CL~09104+4109 are indeed the same cluster, 
it would be about a 10\% coincidence that we happen to see one 
image just before the event started and one during the event. 
This is not a strong argument against identity.

 The time difference of about $3${\hGyr} between these two images
and that of Coma implies that the AGN would most probably have
finished by the epoch of Coma, so would provide no constraint
at all, whether seen or not in the latter.

\subsection{Observational Predictions of the Topological Identity}

\subsubsection{Further Topological Images}

The hypothesised generators imply other images of the
would-be single cluster. The most easily observable of these,
at redshifts low enough that the cluster is known to already
exist, and at which uncertainty in the metric parameters does
not imply too much uncertainty in predicted positions, would
be those at the antipodes (Coma centred)
to RX~J1347-1145 and CL~09104+4109.

These positions are
\begin{eqnarray}
 \rComa - &(\yhat - \rComa) =&  \nonumber \\
 &(0\.40\pm0\.01, &20^h40^m\pm6^m, -40\deg40\arcm\pm20\arcm) \nonumber \\
 \rComa - &(\xhat - \rComa) =& \nonumber \\ 
 &(0\.40\pm0\.01, &01^h57^m\pm4^m, +17\deg25\arcm\pm70\arcm) \nonumber \\
\end{eqnarray}
where $\xhat$ and $\yhat$ are as above, $\rComa$ is the position
of Coma, and the values are $(z,\alpha,\delta)$ (J2000.0).

RX~J203150.4-403656 is a candidate for the former image. 
Spectroscopy to determine its redshift 
and an optical (or X-ray) 
search for a cluster within $\ltapprox 1\.5\deg$ of ($1^h57^m, +17\.5\deg$) 
and within $\delta(z) \ltapprox 0\.01$ of $z=0\.40$ would be needed to
either increase the precision of
or rule out variants on the candidate manifold suggested. In the latter
case, lack of deep ROSAT exposure 
and a foreground Abell cluster make the 
search for a cluster in this region difficult.

\subsubsection{Infall of Sub-cluster(s)}
For the cooling flow in RX~J1347.5-1145 
to be disrupted by subcluster merging, and for subcluster merging to
continue by the time we observe Coma,
the infalling sub-cluster(s) should be visible close 
to RX~J1347.5-1145.  ROSAT pointed observations or 
deep optical imaging over a 20--30$'$ field
around the cluster should be sufficient to detect the subclusters 
if their relative transversal 
velocities are no more than around 1500\k.

\section{Conclusion} \label{s-concl}

The techniques of deducing galaxy transverse velocities and
foreground/background spatial offsets relative to a cluster `central'
galaxy, in the case that a non-trivial topology of the
Universe is known and has $2\rinj\ll 2R_H,$ have been 
presented.

In the case of a 3-manifold for which there are two topological
images of a given galaxy cluster
\begin{list}{(\roman{enumi})}{\usecounter{enumi}}
\item  only separated by a translation (no rotation nor reflection)
and
\item for which the time interval between the two images is 
relatively short (much less than a dynamical time of the cluster);
\end{list}
then 
\begin{list}{(\roman{enumi})}{\usecounter{enumi}}
\item there is at least one common axis to the two sky planes, which 
is simply given by the normal
to the two vectors pointing to the two images; 
\item simple criteria can be chosen to match galaxies between 
the two images;
\item the shifts of corresponding galaxies relative to the cluster centre 
over the time interval imply transverse velocities and
\item if the image separation is closer to 90{\deg} than to 
0{\deg} or 180{\deg}, then the remaining sky offset components
can be used to deduce the other components of the full three-dimensional
relative positions of the galaxies. Spectroscopy would enable the
remaining two velocity components to be estimated.
\end{list}

For a 3-manifold for which two topological images are separated
by a less simple isometry, the present calculation would have to
be generalised.

For the hypothesised identification of the rich clusters 
RX~J1347.5-1145 and CL~09104+4109, scans of the UK SRC Schmidt Southern 
Sky Survey and the Palomar Optical Sky Survey digitised plates were
subjected to a simple version of the criteria suggested for 
matching galaxies between the two images. 

The common sky-plane axes, transverse velocities, 
spatial background/foreground offsets and photometric evolution of
matched galaxies were calculated.
Only one of the transverse velocities is rejected on the grounds of 
being much higher than the velocity dispersion of the would-be
single cluster.
A control sample of randomly orientated axes gives 
about a 20\% chance of inferring a similarly reasonable set of velocities.
The photometric evolution is consistent with the galaxy matches,
so is not useful (in this case at least) for refuting the
original hypothesis.

This method shows how galaxy transverse velocities can be 
estimated, given a knowledge of multiply topologically imaged
clusters by other means. Given the ease of finding reasonable
velocities for a 3-manifold candidate which is in no way anything
more than a candidate, it is clear that this method is not
optimal for the inverse process,
i.e. for refuting a candidate for the topology of the Universe.
It is possible that deep photometry and spectroscopy could enable
such a refutation, though this would at a minimum require a full
scale observing programme.

Regarding the hypothesis adopted, the reader 
is reminded that observations to determine the redshift of 
RX~J203150.4-403656 and to search for a cluster
within $\ltapprox 1\.5\deg$ of ($01^h57^m, +17\.5\deg$) 
and $\delta(z) \ltapprox 0\.01$ of $z=0\.40$ would be useful to
strengthen either (i) arguments against or 
(ii) the precision of the observational 
predictions of the candidate manifold suggested.

This technique of measuring transverse velocities could eventually 
have many other applications. For example, 
to the extent that the generators representing
the topology of the Universe are known precisely enough, quasar
transverse velocities could be measured and be used to improve
corrections to fundamental coordinate reference systems
based on VLBI imaging of quasars (e.g. \citealt{Souch95}).

\section*{acknowledgements}
Special thanks go to the Comit\'e national des astronomes et physiciens 
(CNAP), which inspired this project. Thanks to 
\mbox{Micha{\slash \rule{-0.8ex}{0ex}l}}   
Chodorowski for a careful reading of the manuscript and helpful
suggestions.
This research has been supported by the 
Polish Council for Scientific Research Grant
KBN 2 P03D 008 13 and has benefited from 
the Programme jumelage 16 astronomie 
France/Pologne (CNRS/PAN) of the Minist\`ere de la recherche et
de la technologie (France).
Use has also been made of
the Centre de donn\'ees astronomiques de Strasbourg (CDS) 
of the Observatoire de Strasbourg,
the APMCAT facility at {\em http://www.ast.cam.ac.uk/{\~{}}apmcat} 
and the NASA/IPAC extragalactic database (NED) 
which is operated by the JPL, Caltech, under contract
with the National Aeronautics and Space Administration.

\end{document}